\documentclass[a4paper]{elsarticle}

\usepackage[english]{babel}
\usepackage[utf8x]{inputenc}
\usepackage[T1]{fontenc}

\usepackage{xcolor}

\usepackage[a4paper,top=3cm,bottom=2cm,left=3cm,right=3cm,marginparwidth=1.75cm]{geometry}

\usepackage{graphicx}
\usepackage[colorinlistoftodos]{todonotes}
\usepackage[colorlinks=true, allcolors=blue]{hyperref}
\usepackage{float}
\usepackage{multirow}
\usepackage[normalem]{ulem}
\definecolor{dgreen}{rgb}{0,.6,0}

\usepackage{array}
\newcolumntype{L}[1]{>{\raggedright\let\newline\\\arraybackslash\hspace{0pt}}m{#1}}
\newcolumntype{C}[1]{>{\centering\let\newline\\\arraybackslash\hspace{0pt}}m{#1}}
\newcolumntype{R}[1]{>{\raggedleft\let\newline\\\arraybackslash\hspace{0pt}}m{#1}}
\begin{document}

\begin{frontmatter}
\title{A compression based framework for the detection of anomalies in heterogeneous data sources}

\author[UAM]{Gonzalo de la Torre-Abaitua} \author[UAM]{Luis
  F. Lago-Fern\'andez} \author[CSIC]{David Arroyo \corref{corr}}
\cortext[corr]{Corresponding author: David Arroyo
  (david.arroyo@csic.es).} \address[UAM]{Departamento de
  Ingenier\'{\i}a Inform\'{a}tica, Escuela Polit\'{e}cnica Superior,
  Universidad Aut\'{o}noma de Madrid, 28049 Madrid, Spain} \address[CSIC]{Institute of Physical and Information
  Technologies (ITEFI), Spanish National Research Council
  (CSIC), 28006 Madrid, Spain}

\begin{abstract}  
  Nowadays, information and communications technology systems are
  fundamental assets of our social and economical model, and thus they
  should be properly protected against the malicious activity of
  cybercriminals.  
  Defence mechanisms are generally articulated around tools that trace
  and store information in several ways, the simplest one being the
  generation of plain text files coined as security logs.  This log
  files are usually inspected, in a semi-automatic way, by security
  analysts to detect events that may affect system integrity.
  On this basis, we propose a parameter-free methodology to detect
  security incidents from structured text regardless its nature. We
  use the Normalized Compression Distance to obtain a set of features
  that can be used by a Support Vector Machine to classify events from
  a heterogeneous cybersecurity environment. In specific, we explore
  and validate the application of our methodology in four different
  cybersecurity domains: HTTP anomaly identification, spam detection,
  Domain Generation Algorithms tracking and sentiment analysis. The
  results obtained show the validity and flexibility of our approach
  in different security scenarios with a low configuration burden.
\end{abstract}
\end{frontmatter}
\section{Introduction}
\label{sec:intro}
We live in a complex world with multiple and intricate interactions
among countries, companies and people, which are not always
well-intended.  In our time, those relations are preferentially
conducted by Information and Communication Technologies (ICT)
\cite{/content/publication/9789264026780-en}. As the number of devices
that are connected to the Internet has increased, so has grown the
number of malicious agents that try to get profit from systems
vulnerabilities.  These malicious actors can target governments,
companies or individuals using several kinds of attacks. Malicious
activities range from simple attacks such as parameter tampering, spam
or phishing, to more complex menaces such as botnets, Advanced
Persistent Threats (APTs) or social engineering attacks that leverage
Domain Generation Algorithms (DGAs) and Open Source Intelligence
techniques (OSINT) \cite{enisa18}. In order to help to protect their
networks, systems and information assets, cybersecurity analysts use
different tools, such as Intrusion Detection Systems (IDS), Firewalls
or Antivirus. These tools generate huge amounts of information in the
form of text logs with a blend of normal user interactions and
malicious interactions.  Moreover, the proper implementation of the
security lifecycle is highly dependent on the adequate aggregation of
additional records of activity and other sources of digital
evidence \cite{chuvakin,sabottke2015vulnerability}. Due to the ever increasing size of security related data,
the automatic treatment and analysis of text logs is a key component
in the deployment of adaptive and robust defensive strategies
\cite{Curry2013}.  Acknowledging this fact, it is interesting to have
a single methodology that can be used with these logs to simplify and
help analysts with their duties and decision making.  Our approach
consists of developing parameter-free methods for such a goal
\cite{keogh2004towards}.

As a departing point, we are considering five binary classification
problems in four different cybersecurity related areas that enclose
anomalous versus normal URLs, spam versus non spam messages, normal
versus malicious Internet domains and malicious versus normal messages
in fora or social networks. This is a heterogeneous landscape where
the common nexus is determined by sources of information that can be
treated as text. It can be shown that the use of compression
information techniques comes helpful to process textual data with
different codification systems. Indeed, these techniques have been
successfully used in several domains, such as clustering and
classification of genomic information \cite{ferragina2007compression},
language identification \cite{Cilibrasi2006}, plagiarism detection
\cite{Cilibrasi2005} or URL anomaly detection \cite{Yahalom2008}.

In this work, we explore and validate the use of a previous framework
based on compression information techniques \cite{de2017parameter} in a
wider scenario that encompasses several binary classification problems
in the current cybersecurity context.  These include spam detection,
malicious URL identification, DGA targeting and sentiment analysis in
Twitter and in movie reviews. We use the Normalized Compression
Distance (NCD) \cite{Cilibrasi2005} to extract a set of features that
characterise the textual data associated to each problem, and train a
Support Vector Machine (SVM) that classifies the data using these
features. Our results on the different problems show the generality of
this approach, which shows up as a unified framework for the analysis
of textual information in a general cybersecurity context.

The rest of the
paper is organized as follows: Section
\ref{sec:setting_problem} describes previous works 
on each of the considered domains, Section \ref{sec:methodology}
describes the proposed methodology, Section \ref{sec:Experiments}
explains the details and results of each of the experiments
and Section \ref{sec:conclusions} outlines the conclusions.

\section{Anomaly detection in cybersecurity}
\label{sec:setting_problem}

This section discusses different strategies followed by the
cybersecurity community to tackle some of the threats they
face. Traditional approaches involve the use of anomaly detection
techniques, which have been applied, for example, for virus or
intrusion detection. More recently, with the rise of social networks
and open sources of information, the attack surface has increased
\cite{enisa18}. To overcome these new risks, cybersecurity researchers
have started using Natural Language Processing (NLP) techniques, such
as sentiment analysis. For instance, NLP has been applied for
cyberbulling detection \cite{AutomaticDAHee2015}, malware detection
\cite{killam2016android}, or prediction of cyber-attacks
\cite{HernandezSuarez2018SocialSS}.  In this article we focus on four
representative domains where the analysis of textual information is
also relevant. In the following paragraphs we describe the problems
and the main approaches used to cope with them.

\subsection{HTTP anomaly detection}
\label{subsec:http_anomaly_detection}
Usually, the detection of anomalies in the cybersecurity field is done
using IDS \cite{Garcia-Teodoro2009}.  IDS can be targeted at analysing
either network or host activity. Moreover, we can adopt a static
approach by comparing activity traces to concrete patterns or
signatures, or we can apply a dynamic approach in terms of behaviour analysis. The latter is the one that has focused
most research efforts. In the field of HTTP traffic anomaly detection
several techniques are applied, which can be classified into seven
general groups: statistical models, static analysis, statistical
distances, data mining and pattern analysis, Markov models, machine
learning and knowledge based
\cite{Garcia-Teodoro2009,Chaurasia2016,Bhuyan2014,Hodo,Kruegel,Dong2017}.
Different features of the HTTP packets and the HTTP protocol,
including the text of the URL, are used to model the normal network
behaviour.

\subsection{Spam detection}
\label{subsec:spam_detection}
Spam detection is another area that has been broadly
studied in the cybersecurity context. Several
machine learning techniques have been applied, ranging from
Naïve Bayes to logistic regression or neural networks
\cite{Abu-Nimeh:2007}. As this problem has a big dependence
on text, text analysis and clustering techniques have also been widely
studied, with the Term Frequency (TF) and Term
Frequency Inverse Document Frequency (TFIDF) statistics
\cite{spam_text_cluster} usually applied
for attribute construction, as well
as edit distances \cite{tee2010fpga}. Of particular interest to our work is the
application of compression based techniques
\cite{spam-feature-free,spam_detection_michael}. These
have a good performance
and are quite robust against noise in the channel that eventually could induce an
erroneous classification \cite{Bratko_SFU}.

\subsection{DGA detection}
\label{subsec:dga_detection}
DGA is a technique used by malware to conceal its Command and
Control (C\&C) panel. It consists of the generation of random
domain names and is one of the techniques used by botnets
(groups of infected computers or IoT
devices\footnote{\url{https://www.trendmicro.com/vinfo/us/security/definition/botnet}. Last access 05/07/2019.}) to communicate and hide the real
address of its C\&C.

The main techniques applied to detect DGA
domains analyze the DNS traffic in a specific network. They
use features such as the frequency and the number of domains
that do not resolve (NXDomain) to cluster and identify
malware families \cite{Thomas:2014,Antonakakis:2012}. Another approach
involves the use of the domain name to identify whether it has been
generated by a DGA. This is done by the detection of patterns, and
usually requires a feature extraction step before applying algorithms 
such as Neural Networks \cite{WoodbridgeAAG16,Aashna_dga}
or n-grams \cite{SELVI2019156,Tong_2016}. 

\subsection{Sentiment analysis}
\label{subsec_nlp_sentiment}
Sentiment analysis is a field of text mining that has attracted the attention
of cybersecurity researchers.
Some examples of application of sentiment analysis in the security
context include the identification of threats, radicalism and conflicts in fora 
\cite{Al-Rowaily_BiSAl}, the detection of cybercrime facilitators in
underground economy based on customer feedback \cite{Li-Top_sellers},
and the use of Twitter to anticipate attacks \cite{Ashok_predicting}
and generate alerts \cite{Sudip_CyberTwitter}. The main strategies
consist of a feature selection step using techniques such as TFIDF or
Point-wise Mutual Information (PMI), followed by a sentiment
classification step using either machine learning or lexicon based
approaches \cite{medhat2014sentiment,liu2012survey}. The use of Deep
Learning techniques, including Convolutional Neural Networks
\cite{severyn2015twitter,dos2014deep} and Word Embeddings
\cite{tang2014coooolll}, is also extended.

\section{Methodology}
\label{sec:methodology}
It is worth noting that most of the previously reviewed
techniques are problem specific. Even though all the presented problems
involve some kind of analysis of textual information, the field
lacks a general methodology that faces all these issues from a
common perspective.
This section describes a general mechanism to 
extract a set of numerical attributes that characterize
a text using a compression based metric, such as
the NCD. Given a text $T$, the main idea is to compute the distance
between $T$ and a set of $k$ additional texts $\{g_{1}, g_{2}, ...,
g_{k}\}$, known as {\em attribute generators} (AGs). This provides a
vector of $k$ numbers that represents the text $T$ in a
$k$-dimensional attribute space, and can be used as input for a
classification algorithm such as a SVM \cite{de2017parameter}.

For a dataset consisting of text strings belonging to
two different classes, the detailed
procedure is as follows. We randomly divide the data
into two disjoint groups, $G$ and $I$.
Both groups are balanced, meaning that they contain the same number of
strings from each of the two classes.
The first group, $G$, is used to build the attribute
generators. To do so, the strings in $G$ are packed into a set of
$k$ generator files, each one with strings from one class only. Thus
we have $k/2$ generators for each class. The second group, $I$,
contains the set of strings that are used to train the classifier,
after being characterized by the distances to the
attribute generators. That is, for each string $s_{i}$ in $I$, we
compute the distance between $s_{i}$ and each of the $k$ attribute
generators to obtain the vector:

\begin{equation} \label{eq:Si}
S_{i} = (x_{i1}, x_{i2}, ..., x_{ik})  
\end{equation}

\noindent which is used to characterize the string $s_{i}$. The components $x_{ij}$ of
$S_{i}$ are given by:

\begin{equation} \label{eq:xij}
x_{ij} = D(g_{j}, s_{i})  
\end{equation}

\noindent where $D(g_{j}, s_{i})$ is the normalized conditional
compressed information \cite{Cilibrasi2005} between the
generator $g_{j}$ and the string $s_{i}$:

\begin{equation} \label{eq:dab}
D(g_{j}, s_{i}) = \frac{C(s_{i}|g_{j})}{C(s_{i})} = \frac{C(g_{j}
  \cdot s_{i}) - C(g_{j})}{C(s_{i})}  
\end{equation}

\noindent where $C(x)$ is the compressed size of $x$ and the dot
operator represents concatenation. We use the {\em gzip}
compressor because it has a better speed performance than other
compression algorithms \cite{lzma_benchmark}.  

\begin{figure} [H]
\centering
\includegraphics[width=\textwidth]{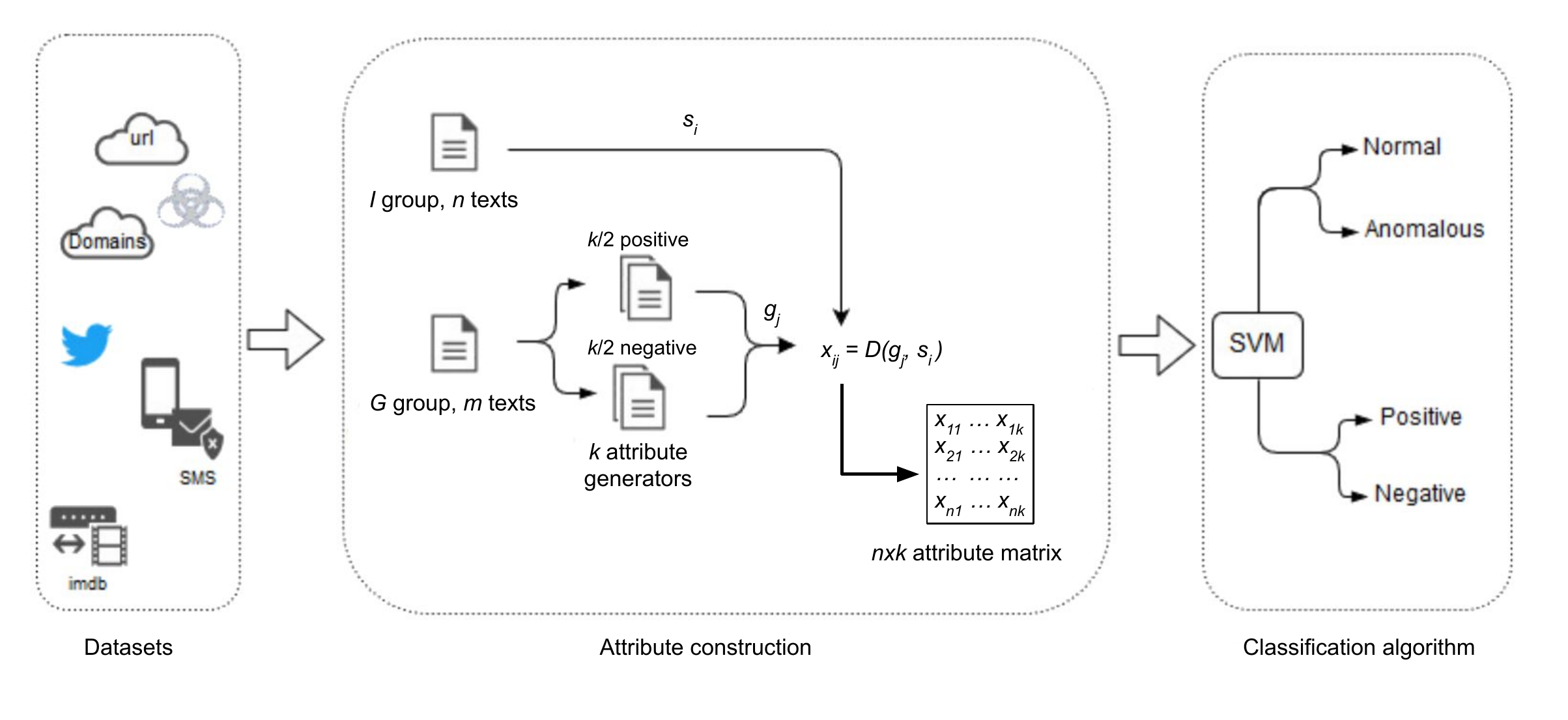}
\caption{End to end representation of the proposed methodology. The set of texts composing the dataset are first divided into two groups, $G$ and $I$. The texts in the $G$ group are further divided into $k$ additional groups, or {\em attribute generators}. Then the distance between each text in $I$, $s_{i}$, and each generator $g_{j}$ is computed in order to obtain a $n \times k$ attribute matrix which is used to train and validate a SVM.}
\label{fig:methodology}
\end{figure}

Once the attribute vectors have been generated for all the strings in
$I$, we use them to train a SVM with a RBF kernel that predicts the
class associated to each string.  We use the {\em scikit-learn} Python 
library \cite{scikit-learn} for the implementation, and measure the
quality of the classifiers by using the accuracy (ACC) and the area
under the receiver operating curve (AUC) metrics. The results
presented in Section \ref{sec:Experiments} are averages over different
partitions of the data into the $G$ and $I$ sets, following the same
validation procedure as in \cite{de2017parameter}. Figure
\ref{fig:methodology} shows the end to end process, from the initial
dataset to the final classification. 

\section{Experiments}
\label{sec:Experiments}
To assess the validity of the proposed methodology, we have performed
experiments in four different and heterogeneous domains. Concretely, we
have explored the detection of malicious HTTP requests
(Section \ref{subsec:Experiment_url}), the identification of spam in SMS messages
(Section \ref{subsec:Experiment_spam}), the detection of DGA domains
(Section \ref{subsec:Experiment_dga}) and the analysis of sentiment both in Twitter
(Section \ref{subsec:Experiment_twitter}) and in movie reviews
(Section \ref{subsec:Experiment_movies}).
All the problems consist of a set of strings belonging to two
different classes, for example normal versus anomalous HTTP requests or
positive versus negative sentiment in tweets. Nevertheless the
characteristics of each dataset are unique, and we observe a high
variability both with respect to the
string length (table \ref{tab:average_len_strings}) and to the string
content (tables \ref{tab:sample_url_strings}, 
\ref{tab:sample_sms_strings}, \ref{tab:sample_dga_normal_domains},
\ref{tab:sample_tweets} and \ref{tab:sample_movies_review}).
Despite this variability, the features extracted following the
presented methodology are able to provide a good description of the
problem data in all the cases, and the classifiers trained on them
obtain state of the art accuracy.
The following subsections describe in detail each of the experiments
carried out. 

\begin{table} [H]
    \caption{\label{tab:average_len_strings} Average (Avg.) and median
      (Med.) of the string length for each of the problems and
      classes.}
\centering
\begin{tabular}{| C{0.8cm} || C{1cm} | C{0.8cm} || C{1cm} | C{0.8cm} || C{1cm} | C{0.8cm} || C{1cm} | C{0.8cm} || C{1cm} | C{0.8cm} |}
\hline 
\multirow{2}{*}{} & \multicolumn{2}{c||}{URL} &
\multicolumn{2}{c||}{DGA} & \multicolumn{2}{c||}{spam} &
\multicolumn{2}{c||}{Twitter} & \multicolumn{2}{c|}{Movie reviews}\\
\cline{2-11} 
& $normal$ & $anom.$ & $normal$ & $DGA$ & $normal$ & $spam$ & $posit.$ & $negat.$ & $posit.$ & $negat.$ \\
\hline
Avg. & $139$ & $152$ & $15$ & $20$ & $72$ & $139$ & $74$ & $75$ & $1325$ & $1295$\\ 
\hline
Med. & $72$ & $95$ & $15$ & $19$ & $53$ & $150$ & $69$ & $70$ & $969$ & $974$\\ 
\hline 
\end{tabular}
\end{table}

\subsection{Experiment 1 - Malicious URL}
\label{subsec:Experiment_url}

This experiment tackles the issue of identifying a malicious HTTP
request only using the related URL string. Usually, this problem has been
faced analysing additional information, such as the URL length, the number
of URL parameters or the parameter values
\cite{torrano2015combining}. Our methodology simplifies the
preprocessing step by considering the raw text of the URL, hence
avoiding any sort of manual attribute construction.

\subsubsection{Data Preparation.}
\label{subsubsec:Experiment_malicious_url_data_preparation}
We use the public $CSIC-2010$ dataset \cite{csic}, which contains
examples of both normal and anomalous HTTP queries. We extract all the
POST requests, remove duplicates and balance the queries. This results
in a total of 9600 queries, 4800 of each class. After this 
preprocessing step, we divide the dataset into the $I$ set, with 1600
randomly chosen queries (800 normal and 800 anomalous) and the $G$
set, with 8000 queries (4000 normal and 4000 anomalous). Some 
examples of normal and anomalous queries are shown in Table 
\ref{tab:sample_url_strings}.

\begin{table} [!htbp]
    \caption{\label{tab:sample_url_strings} Sample strings for $4$ normal
      and $4$ anomalous queries in the $CSIC-2010$ dataset.}
\centering
\begin{tabular}{ l }		
   {\bf Normal queries} \\
   \hline
   \scriptsize \verb|modo=registro&login=beveridg&password=camale%F3nica&nombre=Stefani&apellidos=Gimeno+Cadaveira&|\\
   \scriptsize \verb|  email=morando%40sandrasummer.gy&dni=91059337Z&direccion=C%2F+Bobala+111+4A&ciudad=Mog%E1n&|\\
   \scriptsize \verb|  cp=46293&provincia=Castell%F3n&ntc=8975566519527853&B1=Registrar|\\
   \hline
   \scriptsize \verb|id=1&nombre=Queso+Manchego&precio=39&cantidad=57&B1=A%F1adir+al+carrito|\\
   \hline
   \scriptsize \verb|modo=insertar&precio=5588&B1=Confirmar|\\
   \hline
   \scriptsize \verb|modo=registro&login=ouellett2&password=radicalmente&nombre=Ranquel&apellidos=Orra&|\\
   \scriptsize \verb|  email=hodo%40deltamarina.my&dni=18518539C&direccion=Fructuoso+Alvarez%2C+55+&|\\
   \scriptsize \verb|  ciudad=Bay%E1rcal&cp=17742&provincia=Palencia&ntc=3322562441567993&B1=Registrar|\\
   \hline
   \\
   {\bf Anomalous queries} \\
  \hline  
  \scriptsize \verb|modo=insertar&precioA=900&B1=Pasar+por+caja|\\
  \hline
  \scriptsize \verb|modo=entrar&login=dedie&pwd=M50879RIST44&|\\
  \scriptsize \verb|  remember=bob%40%3CSCRipt%3Ealert%28Paros%29%3C%2FscrIPT%3E.parosproxy.org&B1=Entrar|\\
  \hline
  \scriptsize \verb|modo=registro&login=alix&password=venI%21A&nombreA=Imelda&apellidos=Delb%F3n+Coll&|\\
  \scriptsize \verb|  email=holister%40brunoseguridad.cf&dni=80525673M&direccion=Plza.+Noria+De+La+Huerta+68%2C+&|\\
  \scriptsize \verb|  ciudad=Alcudia+de+Veo&cp=28690&provincia=%C1vila&ntc=6551003767368321&B1=Registrar|\\
  \hline
  \scriptsize \verb|modo=entrar&login=bienek&pwd=cloqu%27e%2Fro&remember=off&B1=Entrar|\\
   \hline  
\end{tabular}
\end{table}

\subsubsection{Results.}
\label{subsubsec:Experiment_malicious_url_results}
We have performed experiments using a number of attributes, $k$,
ranging from 8 to 160. The results can be seen in Table
\ref{tab:results_urls}. The highest accuracy is obtained for $k=80$,
with a $95\%$ of correctly classified HTTP queries. This value is 
similar to other results reported in the literature
\cite{Nguyen2011}. Nevertheless, our approach does not require a
feature selection step and depends on a smaller number of
hyperparameters.

\begin{table} [H]
    \caption{\label{tab:results_urls} Results on the HTTP
      problem. Mean accuracy (Acc.) and area under ROC (AUC) for each
      of the considered $k$ values. The best values are shown in
      boldface.}  
    \centering

\begin{tabular}{c  c  c  c  c}
\hline 
\hline 
$k$ & $C$ & $\gamma$ & Acc. & AUC \\\hline
8 & 1.0 & 100.0 & $0.84 \pm 0.03$ & 0.909 \\ 
16 & 10.0 & 10.0 & $0.88 \pm 0.03$ & 0.946 \\ 
32 & 1.0 & 10.0 & $0.93 \pm 0.02$ & 0.968 \\ 
80 & 1.0 & 10.0 &  $\bf{0.95 \pm 0.02}$ & \bf{0.975} \\ 
160 & 2.0 & 5.0 & $0.95 \pm 0.02$ & 0.974 \\
\hline 
\hline 
\end{tabular}
 
\end{table}

\subsection{Experiment 2 - spam}
\label{subsec:Experiment_spam}
In this experiment we apply the methodology to the problem
of discriminating between legitimate and spam SMS messages. One 
of the main characteristics of this kind of messages is that they
are usually written using a very informal language, with many invented
terms and abbreviations which are not always grammatically
correct. This fact may be a problem for traditional NLP methods based
on lemmatization or parsing \cite{manning2014stanford}.
The method here proposed is however agnostic to the grammar or the
rules followed by the messages, and it can be directly applied to this
problem without any adaptation. The next paragraphs describe the data 
preparation and the results obtained.

\subsubsection{Data Preparation.}
\label{subsubsec:Experiment_spam_data_preparation}
We use the SMS Spam Collection v.1, a public dataset that can be obtained from the authors'  
page\footnote{\url{http://www.dt.fee.unicamp.br/~tiago/smsspamcollection/}
  Last Access 12/05/2019} and also from
Kaggle\footnote{\url{https://www.kaggle.com/uciml/sms-spam-collection-dataset/data}
  Last Access 12/05/2019}. It contains $5574$ SMS messages, $747$ of
them labeled as {\em spam} and the rest, $4827$, labeled as neutral,
or {\em ham}. We balance the dataset by taking all the spam messages
and randomly selecting a sample of $747$ ham messages. The balanced
data are further divided into the $I$ set, with $400$ messages ($200$
ham and $200$ spam), and the $G$ set, with $1094$ messages ($547$ of
each class). Table \ref{tab:sample_sms_strings} shows some examples of
ham and spam messages. 

\begin{table} [!htbp]
    \caption{\label{tab:sample_sms_strings} Sample strings for several
      ham and spam messages in the SMS Spam Collection v.1 dataset.}
\begin{tabular}{ p{15cm} }		
   {\bf Ham SMS} \\
   \hline
   \scriptsize \verb|What you doing?how are you?|\\
   \hline
   \scriptsize \verb|Ok lar... Joking wif u oni...|\\
   \hline
   \scriptsize \verb|Cos i was out shopping wif darren jus now n i called him 2 ask wat present he wan lor. Then he started guessing who|\\ \scriptsize \verb|i was wif n he finally guessed darren lor.|\\
   \hline
   \scriptsize \verb|MY NO. IN LUTON 0125698789 RING ME IF UR AROUND! H*|\\
   \hline
   \scriptsize \verb|dun say so early hor... U c already then say...|\\
   \hline
   \\
   {\bf Spam SMS} \\
  \hline  
  \scriptsize \verb|FreeMsg: Txt: CALL to No: 86888 & claim your reward of 3 hours talk time to use from your phone now! ubscribe6GBP|\\ \scriptsize \verb|mnth inc 3hrs 16 stop?txtStop|\\
  \hline
  \scriptsize \verb|Sunshine Quiz! Win a super Sony DVD recorder if you canname the capital of Australia? Text MQUIZ to 82277. B|\\
  \hline
  \scriptsize \verb|URGENT! Your Mobile No 07808726822 was awarded a L2,000 Bonus Caller Prize on 02/09/03! This is our 2nd attempt to |\\ \scriptsize \verb|contact YOU! Call 0871-872-9758 BOX95QU|\\
   \hline  
\end{tabular}
\end{table}

\subsubsection{Results.}
\label{subsubsec:Experiment_spam_results}
As before, we have performed experiments for $k$ ranging between $8$ and
$160$. A summary of the results may be found in Table
\ref{tab:result_spam_sms}. We observe an increase of performance as
more attribute generators are used, with a maximum of $0.96$ AUC and
$0.904$ accuracy for $k=160$.
These values are slightly worse than the best results reported in the
literature for the same dataset  \cite{Almeida_sms_spam}, but the latter need a more complex
and problem specific preprocessing step which is avoided if using the
proposed methodology, with the subsequent simplification of the
overall process.

\begin{table}[!htbp]
\centering
\caption{\label{tab:result_spam_sms} Results on the spam problem. Mean accuracy (Acc.) and area
      under ROC (AUC) for each of the considered $k$ values. The best values are 
      shown in boldface. Column labeled $k$ shows the number of
      files. Columns $C$ and $\gamma$ show the SVM parameters.}
\begin{tabular}{c c c c c}
\hline
\hline
 $k$ & $C$ & $\gamma$ & Acc. & AUC  \\ 
\hline
 8 &  1.5 & 100 & $0.8005 \pm 0.02$ & $0.85$ \\
 16 & 1.5 & 50  &  $0.827 \pm 0.02$ & $0.89$ \\
 32 & 1.5 & 25  & $0.85 \pm 0.01$ & $0.92$ \\
 80 & 5 & 25 & $0.889 \pm 0.02$ & $0.95$ \\
 160 & 1.5 & 25 & $\bf{0.904 \pm 0.01}$ & $\bf{0.96}$ \\
\hline
\hline
\end{tabular}
\end{table}

\subsection{Experiment 3 - DGAs}
\label{subsec:Experiment_dga}
In the third experiment we apply the methodology to the detection of 
DGAs relying on the domain name only. The main characteristic of this
problem, which makes an important difference with respect to the rest
of considered scenarios, is that the string lengths are significantly
shorter. It is very unlikely that a domain name contains more than 100
characters (see Table \ref{tab:average_len_strings}). In spite of this
fact the proposed approach has been applied with no adaptations, and
the results are quite satisfactory.

\subsubsection{Data preparation.}
\label{subsubsec:Experiment_dga_data_preparation}

We use a dataset where the legitimate domain names are
extracted from the Alexa top one million
list\footnote{\url{https://www.alexa.com/topsites}, last access 
  12/05/2019}, whilst the malicious domains are generated with 11
different malware families, such as zeus, cryptolocker, pushdo or
conficker. The dataset can be downloaded from Andrey Abakumov's GitHub
repository\footnote{\url{https://github.com/andrewaeva/DGA}, last access
12/05/2019}. Raw data contain 1000000 normal
and 800000 malicious DGA domains. After balancing the
classes, we randomly select a subset of $13000$ domains, $6500$ of
each class. We then use $800$ of these as the $I$ set, with the remaining
$5700$ domains used to build the attribute generators ($G$ set).
Table \ref{tab:sample_dga_normal_domains}
shows some examples of both DGA and normal domain names.

\begin{table} [!htbp]
    \caption{\label{tab:sample_dga_normal_domains} Sample strings for
      normal and malicious domains in the DGA dataset.}
\centering
\begin{tabular}{c || c}
\cline{1-2}
\hline
\hline
 Normal & Malicious \\ 
\hline
 cfre.org & ofdhiydrrttpblp.com \\
 fabricadericos.com & puciftnfkplcbhp.net\\
 earthrootgaming.com & tahathil.ru \\
 google.com & thisarmedindependent.com\\
 facebook.com & cgoellwcvwti.com\\
 mail.ru & ufaqzt.cn\\
\hline
\hline
\end{tabular}
\end{table}

\subsubsection{Results.}
\label{subsubsec:Experiment_dga_results}
A summary of our results on the DGA dataset can be seen in Table 
\ref{tab:result_dga}. As in previous experiments, we have carried out
tests with different $k$ values. The classification accuracy increases
with $k$ up to a point where it saturates. The best results are
obtained for $k$=80, with an accuracy of $0.94$ and an AUC of $0.98$.
These values are better than those reported when using traditional
methods, although they can be improved by using deep neural models
such as recurrent neural networks \cite{Lison2017AutomaticDGA}. Note
however that we are using only a small subset of the original data to
train the classifier.

\begin{table}[!htbp]
\centering
\caption{\label{tab:result_dga} Results on the DGA problem. Mean accuracy (Acc.) and area
      under ROC (AUC) for each of the considered $k$ values. The best values are 
      shown in boldface. Column labeled $k$ shows the number of files. $C$ and $\gamma$ are the SVM parameters.}
\begin{tabular}{c c c c c}
\hline
\hline
 $k$ & $C$ & $\gamma$ & Acc. & AUC  \\ 
\hline
 8 & 100 & 25 & $0.903 \pm 0.008$ & $0.96$ \\
 16 & 0.5 & 25 &  $0.918 \pm 0.005$ & $0.97$ \\
 32 & 0.5 & 25 & $0.9305 \pm 0.004$ & $0.97$ \\
 80 & 1   & 25 & $\bf{0.941 \pm 0.004}$ & $\bf{0.98}$ \\
 160 & 5  & 25 & $0.931 \pm 0.005$ & $0.97$ \\
\hline
\hline
\end{tabular}
\end{table}

\subsection{Experiment 4 - Sentiment analysis in Twitter}
\label{subsec:Experiment_twitter}
For this experiment we use the {\em Sentiment140}
dataset\footnote{\url{http://help.sentiment140.com}, last access
  25/01/2019.} described in \cite{twitterSentiment_AlecGo}, 
  which contains a training set with $1600000$ tweets labeled as 
either positive or negative according to their sentiment.
There are $800000$ positive and $800000$ negative tweets, collected
between April and June 25, 2009, and labeled using the
emoticons contained within the message.  Table \ref{tab:sample_tweets} shows a sample of
5 positive and 5 negative tweets extracted from the training set. The
dataset also contains a small test set with $498$ tweets which were
labeled manually, to be used for validation purposes.

\begin{table} [!htbp]
    \caption{\label{tab:sample_tweets} A sample of 5 positive and 5
      negative tweets extracted from the \em{Sentiment140} dataset.}
\centering
\begin{tabular}{ l }		
   {\bf Positive tweets} \\
   \hline
   \scriptsize \verb|happy sunny sunday peeps  xxx|\\ 
   \hline
   \scriptsize \verb|in worrk now chilling out and cleaning the gym  oiss easy!! im supervisor today hahahahaha go me!!!!|\\
   \scriptsize \verb|Clon show!!!!|\\ 
   \hline
   \scriptsize \verb|is craving sun chips  I think im going to go get some now...lol xo|\\ 
   \hline
   \scriptsize \verb|Congratulations! so glad to hear you've had a great weekend at the markets|\\ 
   \hline
   \scriptsize \verb|I just noticed that I use a lot of smiley faces when I talk.  lmfao|\\ 
   \hline  
   \\
   {\bf Negative tweets} \\
   \hline  
   \scriptsize \verb|I think im going to be pulling a late night to finish this|\\ 
   \hline  
   \scriptsize \verb|Im sad to see my aunt in jail  She was arrested for being a very loud drunk lady|\\ 
   \hline  
   \scriptsize \verb|grrr bad hayfever day already|\\ 
   \hline  
   \scriptsize \verb|Also - I am sunburned. I am hurting  Had a good time at meet yesterday, but walked all over creation|\\
   \scriptsize \verb|and now am very tired.|\\ 
   \hline  
   \scriptsize \verb|I'm so bored. No one is talking on MSN, there's nothing to do, and I've got no texts..|\\ 
   \hline  
\end{tabular}
\end{table}

\subsubsection{Data preparation.}
\label{subsubsec:Experiment_twitter_data_preparation}
In our experiments we use only the tweets in the training set.
Prior to our analysis we preprocess the data in order to remove both
duplicated tweets and tweets that appear both as positive and
negative. After this preprocessing stage we obtain a new training set
with $787956$ tweets of each class. The whole text of the messages,
without any further preprocessing, is used to characterize the
tweets.
The $I$ and $G$ sets are built as in previous sections. In particular we use
$50000$ tweets of each class to build $100$ positive and $100$
negative generators with no more than $32$ kB in size each. Every
generator contains on average $424$ tweets, with this number being
slightly higher for the generators containing positive tweets.
The remaining messages are used to train the classifiers after being
characterized by their distance to the generators. In this case, due
to the size of the dataset, we use a SVM classifier with a linear
kernel. The complexity parameter $C$ is tuned by cross-validation on
$10$ folds.

\subsubsection{Results.}
\label{subsubsec:Experiment_twitter_results}
The results are shown in Table \ref{tab:result_twitter}. The accuracies 
in the table are slightly worse than those reported in
\cite{twitterSentiment_AlecGo}, but there 
are two important points to consider.
First, they perform additional
preprocessing of the tweets. In particular, they replaced all usernames
by the new token {\em USERNAME}, they replaced all urls by the keyword
{\em URL}, and they eliminated repeated letters in order correct some
uses of informal language usually present in tweets. In this article
we have decided to omit these steps in order to show the generality of 
our approach.
Second, their results are obtained on the test set, which contains
only $498$ tweets. Results on such a small dataset may be biased.
In fact, when we evaluate our method on these data, we observe a
systematic increase of both the accuracy and the AUC. 

\begin{table}[!htbp]
\centering
\caption{\label{tab:result_twitter} Results on the Twitter problem. Mean accuracy (Acc.) and area
      under ROC (AUC) for each of the considered $k$ values. The best values are 
      shown in boldface.  Column labeled $k$ shows the number of domains in
      each AG file for the corresponding $k$. $C$ is the SVM parameter
      (take into account that in this case we are working with a
      linear kernel).}
\begin{tabular}{c c c c c c}
\hline
\hline
 $k$ & $C$ & Acc. & AUC  \\ 
\hline
 8   & $0.1$   & $0.668 \pm 0.003$        & $ 0.731 \pm 0.003$ \\
 16  & $10.0$  & $0.699 \pm 0.003$        & $ 0.770 \pm 0.003$ \\
 32  & $0.01$  & $0.728 \pm 0.002$        & $ 0.805 \pm 0.002$ \\
 80  & $0.01$  & $0.754 \pm 0.002$        & $ 0.835 \pm 0.003$ \\
 160 & $0.01$  & $\bf{0.767 \pm 0.003}$ & $\bf{ 0.849 \pm 0.002}$ \\
\hline
\hline
\end{tabular}
\end{table}

\subsection{Experiment 5 - Sentiment analysis in movie reviews}
\label{subsec:Experiment_movies}
In this last scenario we tackle a sentiment analysis problem
in movie reviews. It has the particularity that the strings are 
of arbitrary length. Concretely, the average length of the movie
reviews in the dataset is $1395$ characters for the positive reviews
and $1295$ for the negative reviews (see Table
\ref{tab:average_len_strings}). 
This characteristic contrasts with the Twitter problem, where the
string length is limited to $140$ characters.
Besides, the use of language and grammar tends to be not so informal,
and the inclusion of abbreviations and emoticons is not so extended.
These are fundamental differences that motivate the application of the
proposed methodology in this problem. 

\subsubsection{Data preparation.}
\label{subsubsec:Experiment_movies_data_preparation}
We use a public dataset from the 
Stanford NLP
group\footnote{\url{http://ai.stanford.edu/~amaas/data/sentiment/}, 
  last access 12/05/2019}. It contains 50000 movie
reviews extracted from the
  Internet Movie Database
  (IMDb)\footnote{\url{https://www.imdb.com/}, 
  last access 12/05/2019}. Each review is a text string commenting a
  movie, and classified as either
positive or negative. There are 25000 reviews
classified as positive and 25000 classified as
negative. Table \ref{tab:sample_movies_review} contains 
samples of both classes. From these raw data
we use a subset of $5350$ randomly chosen reviews (half positive, half
negative).
The $I$ and $G$ string groups are build as for previous problems.  
The $I$ set contains $1600$ movie reviews ($800$ positive and $800$
negative), and the $G$ set contains the remaining $3750$ texts.

\begin{table} [!htbp]
    \caption{\label{tab:sample_movies_review} Some examples of
      positive and negative movie reviews in the Stanford dataset.}
\centering
\begin{tabular}{ l }		
   {\bf Positive reviews} \\
   \hline
   \scriptsize \verb|I havent seen that movie in 20 or more years but I remember the attack scene with the horses wearing gas-masks|\\ 
   \scriptsize \verb|vividly, this scene ranks way up there with the best of them including the beach scene on Saving private Ryan,|\\ 
   \scriptsize \verb|I recommend it strongly.|\\
   \hline
   \scriptsize \verb|Now this is what I'd call a good horror. With occult/supernatural undertones, this nice low-budget French movie| \\ 
   \scriptsize \verb|caught my attention from the very first scene. This proves you don't need wild FX or lots of gore to make an | \\
   \scriptsize \verb|effective horror movie.|\\
   \hline
   \\
   {\bf Negative reviews} \\
  \hline  
  \scriptsize \verb|The power rangers is definitely the worst television show and completely ridiculous plastic toy line in the| \\ 
  \scriptsize \verb|history of the United States. There is absolutely nothing even remotely entertaining about this completely|\\ 
  \scriptsize \verb|awful television show.|\\
  \hline
  \scriptsize \verb|Some people are saying that this film was "funny". This film is not "funny" at all. Since when is Freddy Krueger|\\
  \scriptsize \verb|supposed to be "funny"? I would call it funnily crap. This film is supposed to be a Horror film, not a comedy.|\\
  \scriptsize \verb|If Freddy had a daughter, wouldn't that information have surfaced like in the first one!? The ending was also|\\
  \scriptsize \verb|just plain stupid and cheesy, exactly like the rest of it.|\\
   \hline  
\end{tabular}
\end{table}

\subsubsection{Results.}
\label{subsubsec:Experiment_movies_results}

This experiment has been carried out using different $k$ values as in
previous sections. In this case the best results are obtained for
$k=160$ ($acc. = 0.86$, $AUC = 0.93$, see Table
\ref{tab:result_movie_review}). Although these results do not improve
the best reported for the same dataset \cite{standford_movies}, they
are quite competivite, even more if we take into account that they
have been obtained on a small subsample of the original data.

\begin{table}[!htbp]
\centering
\caption{\label{tab:result_movie_review} Results on the Movie
  problem. Mean accuracy (Acc.) and area under ROC 
  (AUC) for each of the considered $k$ values. The best
  values are shown in boldface.
  Column labeled $k$ shows
  the number of files. $C$ and $\gamma$ are the SVM
  parameters.}
\begin{tabular}{c c c c c}
\hline
\hline
 $k$ & $C$ & $\gamma$ & Acc. & AUC  \\ 
\hline
 8 & 0.5 & 0.1 & $0.6975 \pm 0.03$ & $0.78$ \\
 16 & 5 & 0.1 & $0.68225 \pm 0.023$ & $0.75$ \\
 32 & 1.5 & 0.1 & $0.74925 \pm 0.01$ & $0.83$ \\
 80 & 20 & 0.1 & $0.7227 \pm 0.01$ & $0.80$ \\
 160 & 0.5 & 0.1 & $\bf{0.8590 \pm 0.012}$ & $\bf{0.93}$ \\
\hline
\hline
\end{tabular}
\end{table}

\section{Conclusions}
\label{sec:conclusions}
Information security calls for a comprehensive deployment of
protection measures and detection controls. Security logs
are the core of such controls, since they enable event
recording and attacks characterisation. Anomaly detection in
security logs is one of the most relevant means to detect
possible malicious activity. However, those security logs
are derived from plenty of different network and information
flow  modalities. Therefore, there exist an urge to adopt
mechanisms to process security information regardless of the
concrete nature of each log \cite{lillis2016current}. In
this vein, we have proposed a methodology that is able to
process heterogeneous security logs using the NCD to extract
features from them, and subsequently training a SVM to
perform a binary classification. 

To test the methodology, we have performed five experiments over four
different domains. For each of the problems, our results are slightly
below the results reported in the literature in terms of
classification accuracy. Nevertheless, we are using the same procedure
to address each of the problems despite their disparate nature. In
other words, we have sacrificed accuracy to adaptability. In addition,
we are using much less data to train the classifiers than other
proposals in the literature, since part of the available data are used
to build the attribute generators. The use of all training
data, following the approach in \cite{delatorre20}, could lead to
further improvement of the presented results.

Finally, our methodology neither needs to perform
a preprocessing step nor to manually construct features from
the data, and the hyper-parameter tuning is minimal. All in
all, our proposal leads to an adequate trade-off between
training dataset size and performance, and it can be
interpreted as a complementary procedure in methodologies
tackling the limitations of the ``no free lunch'' theorem
\cite{wolpert1997no} by the convenient integration of
several anomaly detection methods.

\section*{Acknowledgements}
This work has been supported by the Comunidad de Madrid
(Spain) under the projects CYNAMON (P2018/TCS-4566) and S2017/BMD-3688,
co-financed with FSE and FEDER EU funds, and by Spanish project
MINECO/FEDER TIN2017-84452-R (\url{http://www.mineco.gob.es/}).

\end{document}